\documentclass[12pt]{article}
\usepackage{geometry}
\usepackage[utf8]{inputenc}
\usepackage{enumitem,amssymb}
\usepackage{ragged2e}
\newlist{thematic}{itemize}{8}
\setlist[thematic]{label=$\square$}

\usepackage{graphicx}
\usepackage{wrapfig}
\usepackage{framed}
\usepackage{xcolor}
\usepackage{tablefootnote}
\usepackage{biblatex}

\usepackage{subfig}
\usepackage{pifont}
\newcommand{\cmark}{\ding{51}}%
\newcommand{\xmark}{\ding{56}}%

\begin{document}
\thispagestyle{empty}
\raggedright
\large
Roman CCS White Paper \linebreak

\textbf{The Galactic Center with Roman} \linebreak
\normalsize

\textbf{Roman Core Community Survey}: Galactic Bulge Time Domain Survey
\linebreak

\noindent \textbf{Scientific Categories:} \hfill Supermassive black holes and active galaxies \linebreak
\hspace*{\fill} Stellar populations and the interstellar medium \linebreak
\hspace*{\fill} Stellar physics and stellar types \linebreak

\noindent \textbf{Additional Keywords:} \hfill Galactic center, Star clusters \linebreak
\hspace*{\fill} Exoplanets and exoplanet formation
\linebreak
\hspace*{\fill} Astronomical simulations

\vspace{-0.2cm}
\textbf{Submitting Author:}

Name: Sean K. Terry
 \linebreak						
Institution: University of California, Berkeley
 \linebreak
Email: sean.terry@berkeley.edu
 \linebreak
 
\textbf{Co-authors:} Matthew W. Hosek Jr. (UCLA), Jessica R. Lu (UC Berkeley), Casey Lam (UC Berkeley), Natasha Abrams (UC Berkeley), Arash Bahramian (Curtin U.), Richard Barry (NASA/GSFC), Jean-Phillipe Beaulieu (IAP/UTAS), Aparna Bhattacharya (UMD), Devin Chu (UCLA), Anna Ciurlo (UCLA), Will Clarkson (U. of Michigan), Tuan Do (UCLA), Kareem El-Badry (Harvard), Ryan Felton (NASA/Ames), Matthew Freeman (UC Berkeley), Abhimat Gautam (UCLA), Andrea Ghez (UCLA), Daniel Huber (U. Hawaii), Jason Hunt (Flatiron/CCA), Macy Huston (PSU), Tharindu Jayasinghe (UC Berkeley), Naoki Koshimoto (Osaka U.), Madeline Lucey (UT Austin), Florian Peißker (U. Cologne), Anna Pusack (UC Berkeley), Clément Ranc (IAP), Dominick Rowan (OSU), Robyn Sanderson (U. of Pennsylvania), Rainer Schödel (IAA-CSIC), Richard G. Spencer (NIH), Rachel Street (LCO), Daisuke Suzuki (Osaka U.), Aikaterini Vandorou (UMD)
\linebreak
\vspace{-0.05cm}

\textbf{Abstract:} 
We advocate for a Galactic center (GC) field to be added to the Galactic Bulge Time Domain Survey (GBTDS). 
The new field would yield high-cadence photometric and astrometric measurements 
of an unprecedented $\sim$3.3 million stars toward the GC.
This would enable a wide range of science cases, such as
finding star-compact object binaries that may ultimately merge as LISA-detectable gravitational wave sources, 
constraining the mass function of stars and compact objects in different environments,
detecting populations of microlensing and transiting exoplanets, 
studying stellar flares and variability in young and old stars, 
and monitoring accretion onto the central supermassive black hole.
In addition, high-precision proper motions and parallaxes would open a new window into 
the large-scale dynamics of stellar populations at the GC,
yielding insights into the formation and evolution of galactic nuclei and their co-evolution with the growth of the supermassive black hole. 
We discuss the possible trade-offs between the notional GBTDS and the addition of a GC field with either an optimal or minimal cadence. 
Ultimately, the addition of a GC field to the GBTDS would dramatically increase the science return of Roman and provide a legacy dataset to study the mid-plane and innermost regions of our Galaxy.

\clearpage
\justifying
\pagenumbering{arabic}
\section*{Summary of Recommendations \& Science:}

In order to conduct one of the deepest wide-field surveys of the Galactic Center (GC), we recommend:
\begin{itemize}
    \item One Roman WFI field positioned at the GC with an identical observing strategy as the notional GBTDS.
\end{itemize}
This will maximize GC-related science and enable a wealth of high-precision, wide-field studies previously unreachable in this region. These include:
\begin{itemize}
    \item A high cadence survey along the GC sight-line that will measure many transient phenomena including young stellar flares, short/long-period variables, star-compact object ellipsoidal variables, near-IR accretion flares from LMXBs, microlensing planets/brown dwarfs/compact objects, transiting planets, and more.
     \item Large scale formation and dynamical evolution of stellar populations in the environment surrounding the supermassive black hole (SMBH) SgrA*.
    \item Long-term, high cadence monitoring of SgrA* events (accretion, close approaches, tidal disruption, lensing, and more).
\end{itemize}

\medskip
\section{Scientific Motivation}

The GC is a unique environment that allows us to study a SMBH 
and the surrounding stellar populations in a resolved manner.
The region is primarily dominated by two old ($\sim$5 -- 10 Gyr) populations: 
the Nuclear Star Cluster (NSC), which
is the densest and most massive star cluster
in the galaxy \cite[][]{Schodel:2014zl},
and the Nuclear Stellar Disk (NSD), 
which extends along the Galactic plane to R $\sim$150 pc \cite{Sormani:2022lz}. 
However, the GC is also a zone of active star formation, 
and hosts three of the most massive young star clusters ($\lesssim$ 10 Myr)
in the MW:
the Young Nuclear Cluster (YNC), the Arches cluster, 
and the Quintuplet cluster \cite[e.g.][]{Serabyn:1998hw, Figer:1999px, Paumard:2006sh, Lu:2013wo, Stostad:2015gr}. 
As similar structures are commonly found in other galaxies,
the GC serves as a valuable template for understanding how 
galactic nuclei form and evolve. In addition to the GC itself, this sight line also probes the MW's inner thin disk with young and metal-rich populations relative to the thick disk and bulge located at the slightly higher galactic latitudes of the GBTDS. We aim to describe the motivation for adding \textit{Roman} observations of the GC with either the same observing schedule as the notional GBTDS or at a lower cadence.

\section{The Survey, Science Cases, and Metrics} \label{sec:science-cases}
\begin{figure}[!h]
  \hspace{-1.5cm}
  \subfloat{\includegraphics[width=0.6\textwidth]{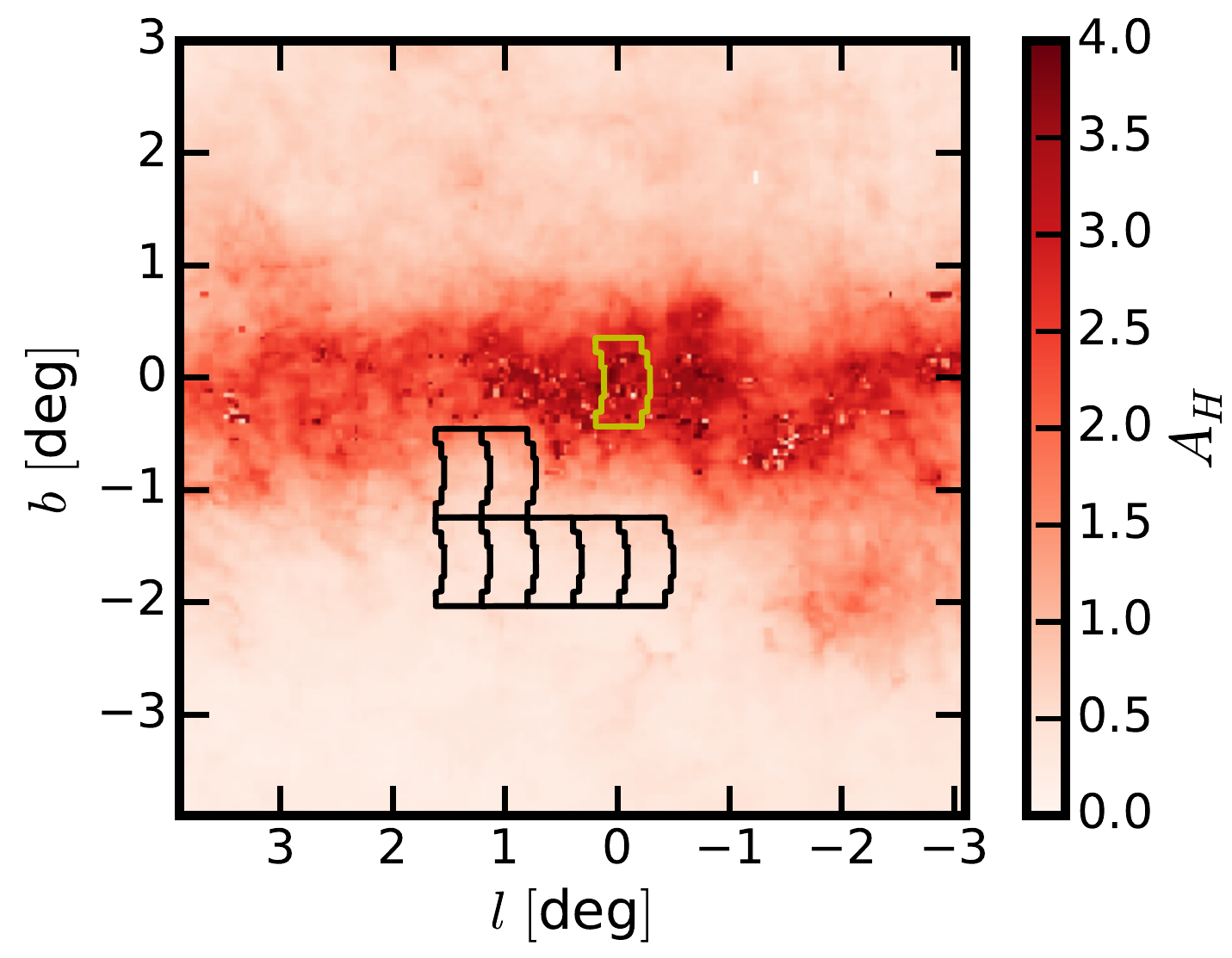}}
  \subfloat{\includegraphics[width=0.56\textwidth]{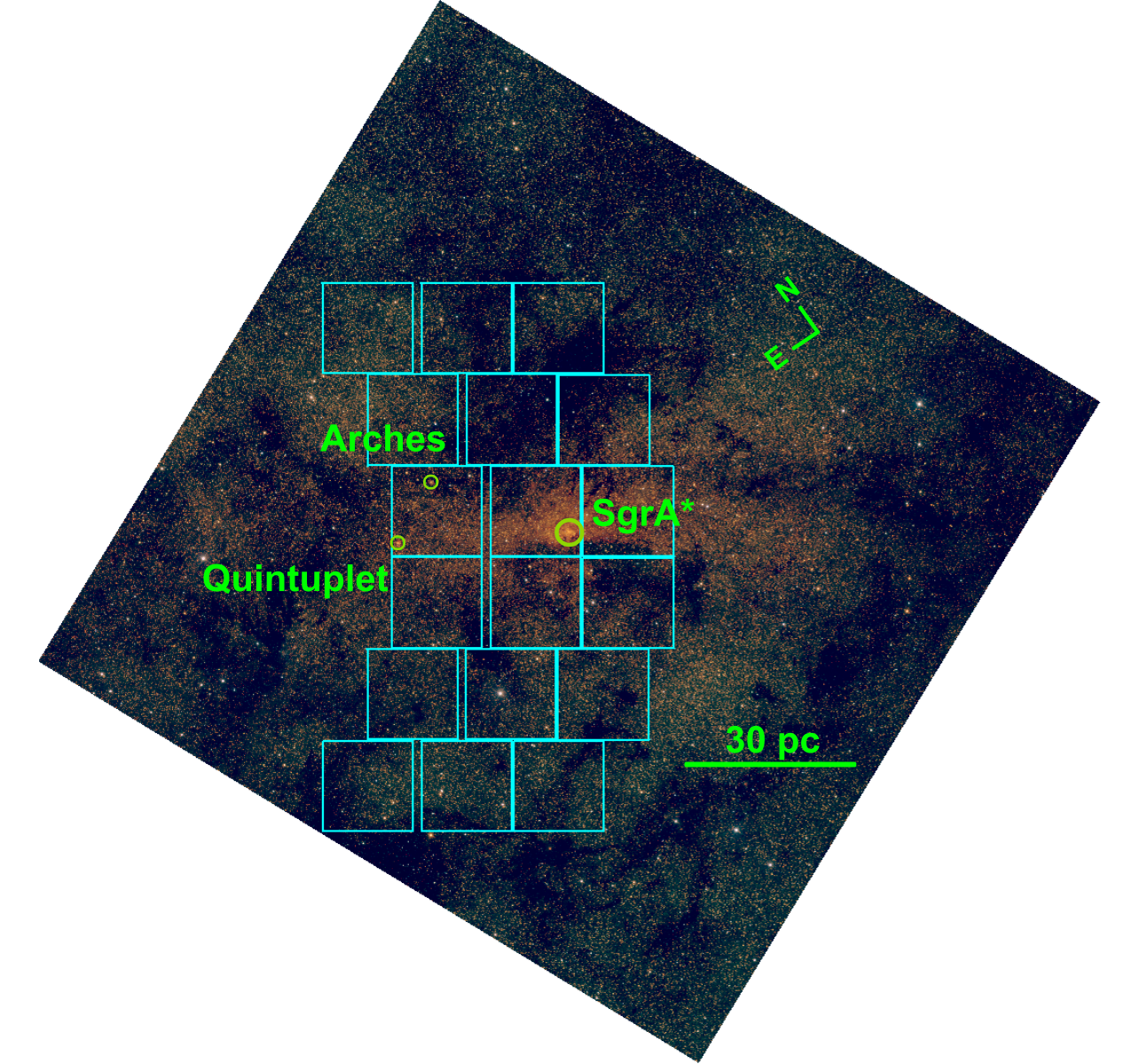}}
  \caption{\footnotesize \textit{Left}: The notional Roman GBTDS fields from \cite{penny:2019a} (black outlines) overlaid on an $H$-band extinction map from \cite{gonzalez:2012a}. The additional \textit{Roman-GC} field described in this paper is shown in yellow outline. \textit{Right}: Composite image (in J, H, Ks bands) of the GC region from 2MASS. The full 0.281 deg$^2$ Roman field covers the central supermassive black hole SgrA*, nuclear star cluster (NSC), young nuclear cluster (YNC), nuclear stellar disk (NSD), Arches, and Quintuplet clusters in one pointing.} \label{fig:combo_map_with_chips}
\end{figure}

The Nancy Grace Roman Space Telescope (\textit{Roman}) will conduct one of the deepest ever surveys toward the central region of the Milky Way (MW) during its Galactic Bulge Time Domain Survey (GBTDS). This Core Community Survey (CCS) will provide multi-filter high resolution imaging of a ${\sim}$2 deg$^2$ patch of sky at approximately $-0.5 < l < 1.5$ and $-2.0 < b < -0.5$ (see Figure \ref{fig:combo_map_with_chips}) for over four years. The Galactic center (GC) lies just ${\sim}1.4$ deg away from the notional GBTDS fields (Figure \ref{fig:combo_map_with_chips}).

We present here an `optimal' and `minimal' observing strategy for the recommended \textit{Roman} GBTDS + GC survey (hereafter \textit{Roman GB+GC}). Both strategies include adding an eighth field at the GC pointing (e.g. 7 GBTDS + 1 GC field; left panel of Figure \ref{fig:combo_map_with_chips}). The optimal strategy involves observing the GC field at the same cadence, exposure times, and filters as the notional GBTDS fields. The minimal strategy involves observing the GC field at a lower cadence (e.g. every 12 hours or more) in one wide passband filter (F146). Both of these strategies would measure a myriad of time-domain phenomena through photometric and astrometric monitoring.

\textbf{A \textit{Roman GB+GC} survey would uniquely probe the transient populations in the Galactic mid-plane and center}. Survey results could include a census of star-compact object binary systems both in the NSC and in the foreground (\ref{sec:star-compact-binaries}). The NSC alone is expected to host many of these systems from star formation and dynamical predictions (\cite{miralda:2000a}; estimated yield of ${\sim}$ 250). Further, we can conduct a census of transiting exoplanets in the innermost Galactic mid-plane (\ref{sec:transits}). The microlensing event rate is at least an order of magnitude larger in the mid-plane \cite{navarro:2017a}, thus the addition of a GC field could add a substantial amount of planetary and compact object detections (${\sim}$hundreds total) to the expected microlensing yield for the GBTDS \cite{johnson:2020a, penny:2019a, sajadian:2023a} and enable us to study how the planet, star, and compact object mass function changes in the Galactic mid plane and center (\ref{sec:microlensing}). Lastly, we can study the occurrence rate of young stellar flares, near-IR counterparts to LMXBs, and accretion events and near-IR flares from SgrA* (\ref{sec:sgra*-monitoring}).
Superb information on stellar variability will provide unique constraints on the stellar population and formation history of the GC. For example, the number density of RR Lyraes will provide us with 
a tight constraint on the presence and distribution of an old metal-poor population in this region.

\textbf{The recommended survey would also enable revolutionary investigations of the 
formation and evolution of its stellar populations through long-term proper motion and parallax measurements}.
For example: how did the NSC and NSD form, and what
is their relationship to each other and to the Galactic bulge at large (\ref{sec:nsc_nsd})?
What is the distribution of mass, and thus gravitational potential of the region (\ref{sec:mass})?
What is the lifecycle, internal kinematics, and stellar Initial Mass Function (IMF) 
of young clusters in the GC environment (\ref{sec:clusters})? 
The outstanding spatial coverage and astrometric performance
offered by Roman makes it uniquely suited
to conduct a large-scale kinematic survey of the region to address these questions.
In the case of the optimal survey, 
it will be possible to make parallax-based measurements of 
stellar distances for bright stars towards the GC (\ref{sec:astrometry}). 



\begin{figure}[!h]
  \includegraphics[width=\textwidth]{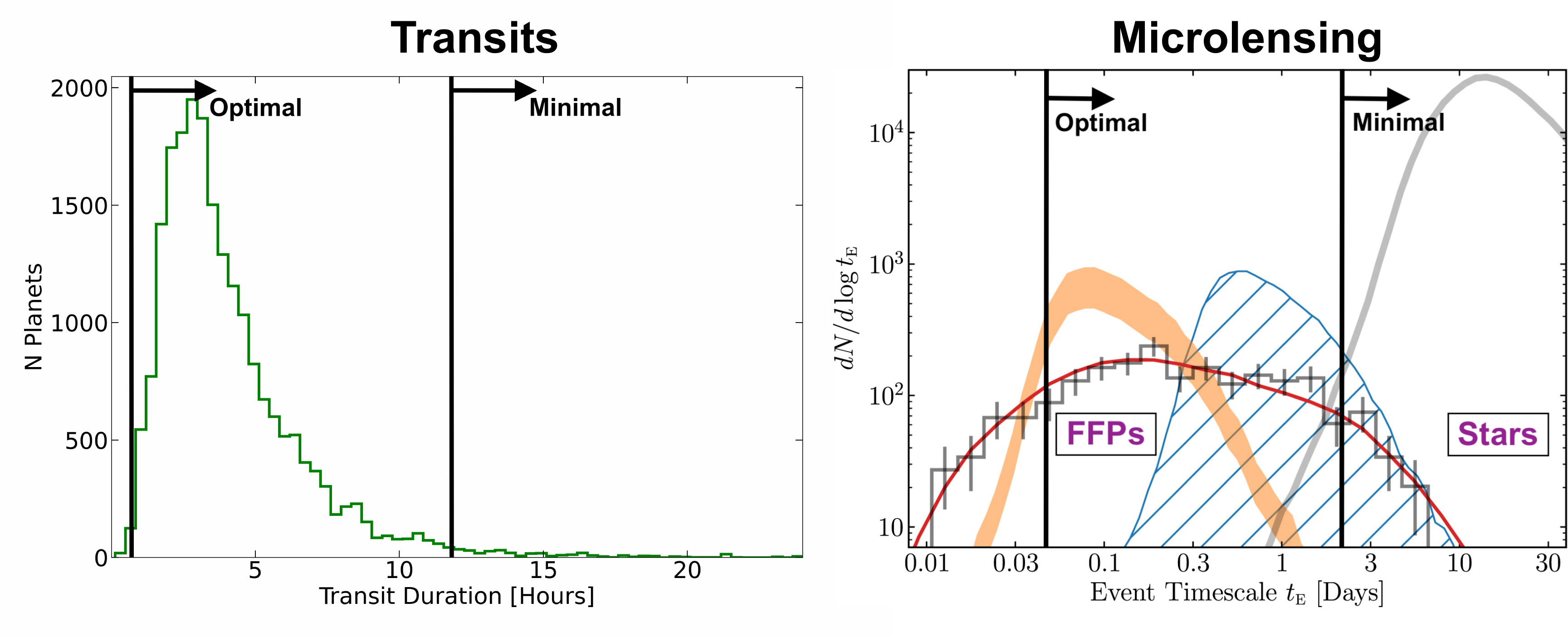}
  \caption{\footnotesize \textit{Left panel}: The distribution of transit duration for confirmed and candidate transiting exoplanets from the NASA Exoplanet Archive$^3$. The minimal and optimal thresholds (black vertical bars) are shown with the cadence of 15 minutes and 12 hours, respectively. The minimal strategy shows that \textit{Roman} will measure at best one data point during transits for long period (transit duration) planets, while the optimal strategy will measure many data points during transit for nearly the entire population. Stacking and phase folding transit light curves will benefit both strategies. \textit{Right panel}: The Einstein timescale distribution for predicted FFPs and low-mass stars, adapted from \cite{johnson:2020a}. The $t_E$ signal from stars is shown as a thick gray line. The FFP distributions (orange, red, blue-hatched) are derived using three different mass functions from \cite{cassan:2012a} and \cite{mroz:2017a}. The minimal and optimal thresholds shown here include a requirement that \textit{Roman} acquires at least five data points during the microlensing event. The optimal strategy here will aid in constraining the true mass function of FFPs, while the minimal strategy will miss a significant fraction of FFPs.}
  \label{fig:transit-ulens-timescales}
\end{figure}

\footnotetext{https://exoplanetarchive.ipac.caltech.edu/}

\subsection{Depth, Photometric and Astrometric Requirements}
\label{sec:req}
\begin{figure}[!h]
  \centering
  \includegraphics[width=0.7\textwidth]{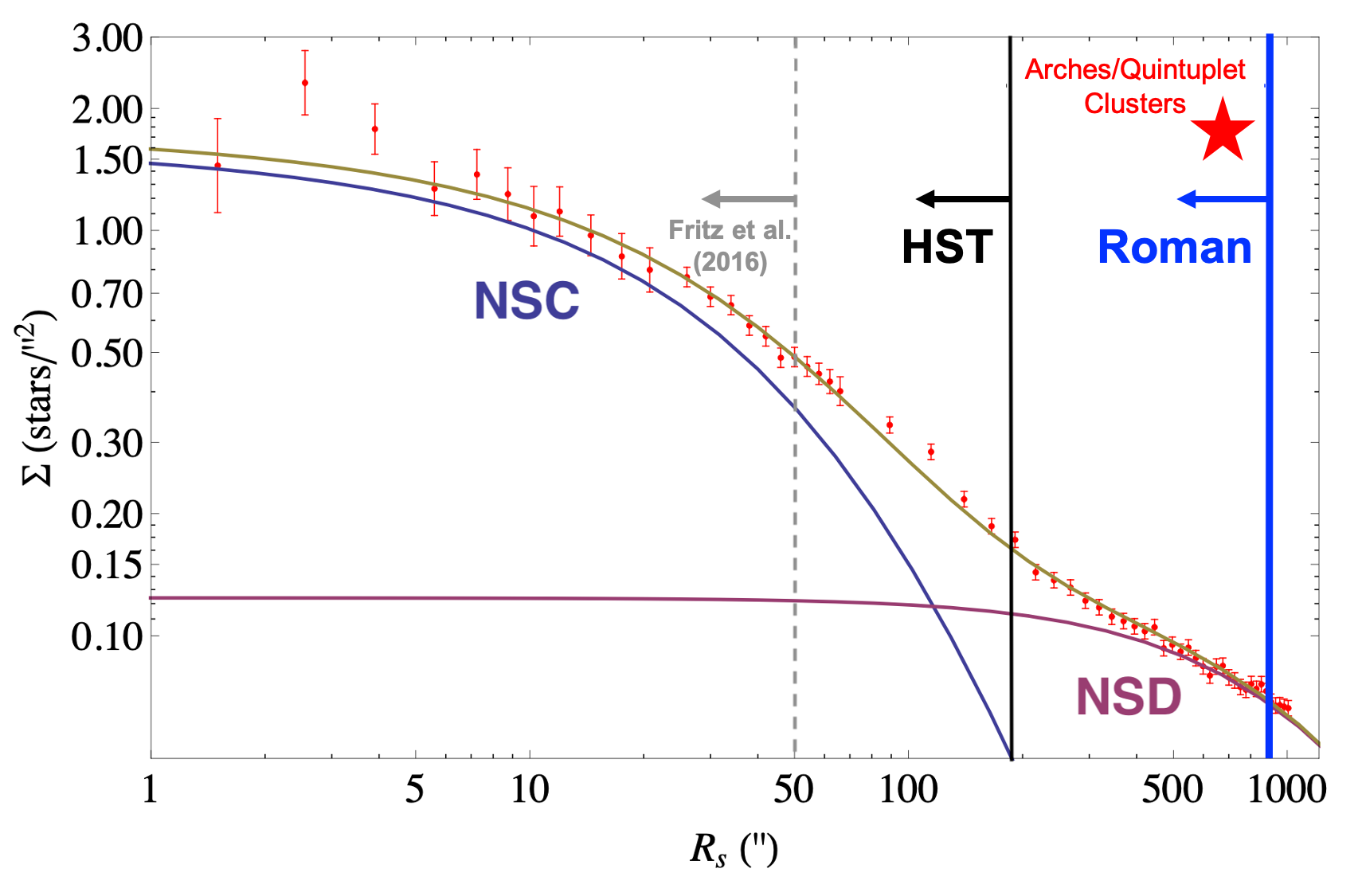}
  \caption{\footnotesize The observed stellar density profile of the GC showing the NSC and NSD, adopted from \cite{Chatzopoulos:2015lq}. The \textit{Roman GB+GC} survey would provide stellar proper motions with at least 50\% azimuthal coverage well into the NSD (blue line), covering an area $\geq$7x larger than previous proper motion surveys (e.g. \cite{Fritz:2016zr} and HST, grey and black lines, respectively). The field would also include the Arches and Quintuplet clusters (red star).} 
  \label{fig:nsc_nsd}
\end{figure}

\
We estimate that \textbf{the GC field will obtain photometric and astrometric measurements of $\sim$3.3 million stars 
over an area at least $\sim$7x larger than previous proper motion studies of the GC} (Fig. \ref{fig:nsc_nsd}).
The depth of the survey (F146 $\sim$ 23$-$24 mag) is set by
the estimated stellar confusion limit for \textit{Roman} based on \textit{HST} WFC3-IR images 
in the GC region \cite[][]{Hosek:2015cs, Rui:2019ae}.
For an Arches-like cluster, this depth corresponds to a 0.5 -- 0.7 M$_{\odot}$ pre-main sequence star. For an old NSC-like population, this depth reaches  0.8 -- 1.1 M$_{\odot}$ stars near the main-sequence turn-off.



The primary driver for the minimal and optimal 
strategies is the detection
and characterization of time-domain phenomena at and along the sight-line to the GC.
The estimated photometric precision achieved by both observing strategies is $0.03 - 0.06$ mag in a single F146 image. 
This precision allows for very subtle light curve features to be detected in microlensing and transiting science cases, 
as well as ellipsoidal variations due to star-compact object binaries, flares from both young stellar objects (YSO) and the near-IR counterpart to SgrA*. 
The range of transients the survey is sensitive to is determined by the cadence of the observations.
The optimal strategy allows for full characterization of hour long flares (SgrA*, \ref{sec:sgra*-monitoring}), planets with transit durations as low as two hours, and short timescale light curve perturbations from low-mass planets \cite{penny:2019a}.
As described in Table \ref{tab:summary}, Figure \ref{fig:transit-ulens-timescales}, and Appendix A, the minimal strategy loses sensitivity to SgrA* flares, the shortest period transiting planets and eclipsing binaries, as well as the lower-mass regime of 
 bound planets and FFPs from microlensing.

Both the minimal and optimal survey strategies will 
produce high-precision stellar proper motions
due to the high cadence of the observations.
The minimal strategy will achieve precisions 
of $\sim$25 $\mu$as/yr ($\sim$1 km/s), 
which enables investigations of 
kinematic sub-structures within the NSC and NSD 
as well as the velocity dispersion profiles of 
the young clusters.
\textbf{With the optimal survey, it becomes possible 
to measure stellar distances at the GC to within $\pm$200 pc via 
parallax for the brighter stars 
in the survey (F146 $\lesssim$ 19 mag) for the first time\footnote{This assumes that the observing seasons will alternate between Fall and Spring windows for the GC.}}.
This distance resolution makes it possible to 
separate stars in the NSC + NSD from those
in the Galactic bulge.

\section{Discussion and Conclusions} \label{sec:conclusion}
We recommend that a GC field be added to the notional GBTDS. This \textit{Roman GB+GC} survey will produce long-baseline wide-field monitoring of SgrA*, the surrounding GC region, and Galactic mid-plane populations along the sight-line. A full Roman field on the GC at the GBTDS cadence will enable detection and characterization of many transient phenomena, from young stellar flares and short/long period variables, to transiting planets in the innermost mid-plane. We have also shown that this survey will gather a census of star-compact object systems that are expected to be hosted by the NSC and surrounding environment. Monitoring of the central SMBH and its local effects will add to the rich, decades-long history of high-precision studies of this object and the population of stars orbiting it \cite{ghez:2008a, genzel:2010a}. Additionally, the proper motions produced by the survey will allow for comprehensive kinematic modeling of the NSC and NSD, constraining their formation and evolution, and make the first measurements of the tidal tails and velocity dispersion profiles of the Arches and Quintuplet clusters, revealing the lifecycle and IMF of young clusters near the GC.

We have demonstrated that a minimal observing strategy of twice a day observations on the GC, taken with a single wide band filter (e.g. F146), enables many of the presented science investigations to be achieved (Table \ref{tab:summary}, Appendix A, B). However, an optimal observing strategy will maximize the scientific return for many of the transient populations that rely on high-cadence photometric monitoring (transiting planets, eclipsing and star-compact object binaries, short duration flares, microlensing planets).

Given the significant number of compelling science cases for a \textit{Roman GB+GC} survey, along with the relative ease of visiting the GC during the notional GBTDS, we advocate that the community survey committee consider this strategy. The addition of this GC survey at the GBTDS cadence will increase the scientific return of the mission as a whole. The exact trade-offs between the cadence, number of fields, and detection yields for the science cases (including microlensing planets) should be explored with a more detailed analysis.\\

\begin{table}[!h]
\centering
\caption{The GC Observing Strategy} 
\label{tab:observing-strat}
\begin{tabular}{lcc}
\hline
        Parameter &  Minimal &  Optimal \\\hline\hline
        Pointing & SgrA$^*$ & SgrA$^*$ \\
        Cadence & \textbf{${\gtrsim}$12 hours} & \textbf{Same as GBTDS} \\
        Depth & 23 mag $<$ F146 $<$ 24 mag & 23 mag $<$ F146 $<$ 24 mag\\
        Total seasons & Same as GBTDS & Same as GBTDS \\
        Total Survey Area & 0.281 deg$^2$ & 0.281 deg$^2$ \\
        Specific filters & \textbf{F146 only} & \textbf{Same as GBTDS} \\
        Total Filters & \textbf{one} & \textbf{Same as GBTDS} \\
        Subpixel Dithers & \cite{sanderson:2019a} & \cite{sanderson:2019a} \\
        Large gap dithers & \cite{sanderson:2019a} & \cite{sanderson:2019a} \\
        Ground-based coverage & Nightly w/ PRIME\tablefootnote{http://www-ir.ess.sci.osaka-u.ac.jp/prime/index.html} & Nightly w/ PRIME$^2$ \\
        Photometric precision & 0.03 -- 0.06 mag & 0.03 -- 0.06 mag  \\
        Astrometric precision & 1 -- 1.6 mas & 1 -- 1.6 mas \\
        Proper Motion precision & \textbf{15 -- 25 $\mu$as/yr} & \textbf{2.5 -- 3.5 $\mu$as/yr} \\
        \hline
\end{tabular}
\\
\vspace{0.2cm}
{\raggedright \footnotesize{\textbf{Note}. Photometric and astrometric precisions are estimated for a single F146 image, while the proper motion precision is estimated for the completed survey. All precisions correspond to the values achieved at the reported F146 depth. Magnitudes are given in the VEGAMAG photometric system.}\par}
\end{table}


\begin{table}[!h]
\centering
\caption{Summary of Science Cases, Achievable by Minimal or Optimal Strategy} 
\label{tab:summary}
\begin{tabular}{lccc}
\hline
        Science Case & Section & Minimal & Optimal  \\\hline\hline
        Star-Compact Object Binaries & \ref{sec:star-compact-binaries} & \xmark/\cmark & \cmark \\
        Transiting Exoplanets and EB's & \ref{sec:transits} & \xmark & \cmark \\
        Microlensing & \ref{sec:microlensing} & \xmark/\cmark & \cmark  \\
        Monitoring SgrA* & \ref{sec:sgra*-monitoring} & \xmark & \cmark \\
        The NSC/NSD & \ref{sec:nsc_nsd} & \cmark & \cmark \\
        Mass Distribution of the GC & \ref{sec:mass} & \cmark & \cmark  \\
        Lifecycle, kinematics, IMF of GC clusters & \ref{sec:clusters} & \cmark & \cmark  \\
        Parallaxes at the GC (F146 $\lesssim$ 19 mag) & \ref{sec:astrometry} & \xmark & \cmark \\
        \hline
\end{tabular}
\\
\vspace{0.2cm}
{\raggedright \footnotesize{\textbf{Note}. For clarity, \cmark means that the given science case can be achieved with the specified strategy, as described in the Appendices. \xmark\, means the science case cannot be achieved. The mixed \xmark/\cmark means the given science case will have diminished sensitivity, but can be partially achieved.}\par}
\end{table}


\appendix
\section{Transients Toward the GC}
\label{subsec:transients}

\subsection{Star-Compact Object Binary Systems} \label{sec:star-compact-binaries}

Regardless of the minimal or optimal observing strategy, \textbf{this \textit{Roman GB+GC} program will gather the largest census of star-compact object binary systems from a single observatory to-date}. For future considerations, these systems can also be precursors to the mergers that are currently being detected by LIGO, as well as those systems that LISA expects to detect via gravitational waves. These objects (sometimes referred to as detached systems) can be characterized through continuous monitoring and photometric light curve modeling of temporal phenomena: ellipsoidal variations, Doppler beaming, and self-lensing (e.g. Masuda \& Hotokezaka 2019). 

We highlight ellipsoidal variables in the NSC as an example of the power of a \textit{Roman GB+GC} program for better understanding binary systems. 
Dynamicists predict that the NSC and surrounding region should host many of these systems from mass segregation and exchange interactions occurring among these binaries. With the minimal \textit{Roman-GC} observing strategy described in Table 1, we estimate ${\sim}275$ candidate detections with the minimal strategy, or ${\sim}350$ candidate detections with the optimal strategy during the full mission. The optimal survey adds ${\sim}75$ short period systems that are uniquely interesting for constraining binary merger models. Furthermore, the additional temporal coverage provided by the optimal survey will help constrain the true period distribution for these systems, which is unknown at present. Confirming these candidates as true star-compact object binaries would require spectroscopic followup.  


\subsection{Probing Binary Stars and Planets with Transit Photometry} \label{sec:transits}
Additionally, with the optimal observing strategy and expected photometric precision of $0.03 - 0.06$ mag (Table \ref{tab:observing-strat}), a high-cadence \textit{Roman} GC field will detect thousands of eclipsing binaries and transiting exoplanets. A simple extrapolation from recent \textit{Roman} transiting exoplanet predictions \cite{tamburo:2023a, wilson:2023a} (adjusting for higher surface density $+$ extinction) gives between ${\sim}8,500$ and ${\sim}28,000$ transiting exoplanets toward the GC. Most of these planets will be giants ($R_p > 4R_{\oplus}$) on close-in orbits ($a < 0.3$ AU), however small planets ($R_p < 4R_{\oplus}$) could make up $5-10\%$ of the total yield. Regardless of the sizes, \textbf{an exciting new window will be opened on the population of transiting planets in the innermost mid-plane of the MW}. 

As shown on the left panel of Figure \ref{fig:transit-ulens-timescales}, the minimal GC observing strategy would miss nearly all of the transiting exoplanet detections. This leaves mostly long period eclipsing binaries as the only reachable populations related to this science case.

\subsection{The Mass Function of Planets, Stars, and Compact Objects via Microlensing in the Mid-plane} \label{sec:microlensing}
The microlensing event rate is expected to increase significantly at lower galactic latitudes. Indeed, a recent study of microlensing events found in the VISTA Variables in the Vía Láctea (VVV) Survey shows the microlensing optical depth rising continuously up to the GC, despite the increased extinction \cite{navarro:2017a}. This large survey mapped the GC region and surrounding galactic disk with near-IR photometry ($K_s$ band) at approximately nightly cadence. This previous study represents an important demonstration that an effective near-IR survey in this field can be profitable and detect microlensing objects with timescales ranging from $t_E\, {\lesssim} \,1.5$ days (e.g. FFPs) to $t_E\, {\gtrsim}\, 100$ days (e.g. stellar-mass black holes.

Using predictions from \cite{penny:2019a, johnson:2020a, sajadian:2023a} [adjusted for higher event rate $+$ extinction], we estimate a ${\sim}$five year \textbf{GC survey (at the optimal cadence) may detect as many as ${\sim}25,000$ microlensing events}. Of these, we estimate ${\sim}25$ FFPs, ${\sim}33$ bound Earth-mass planets, and at least 180 planets (total bound + FFPs). Additionally, we estimate between ${\sim}35$ and ${\sim}75$ isolated black holes and neutron stars may be detectable with this survey. These predicted yields depend strongly on the average timescale ($t_E$) of microlensing events toward the bulge, which may be shorter than the timescales in typically observed microlensing fields. Further, the exact number of compact object lenses that will be characterizable (e.g. reliable measurement of mass and velocity) will depend on the cadence and time baseline between seasons. See the black hole microlensing and Roman-Rubin synergies CCS white papers for more details \cite{Lam:2023a, street:2023a}. The minimal observing strategy presented here would preferentially lose the lowest-mass bound planets and FFPs (see Figure \ref{fig:transit-ulens-timescales}), as low-mass binary lens planetary perturbations and FFPs give the shortest duration light curve features \cite{johnson:2020a, penny:2019a}. However, we still expect ${\sim}$few hundred higher-mass planet detections with the minimal observing strategy, which will aid in constraining parts of the mass function for stars and planets with microlensing \cite{wegg:2017a}.

Given the much higher and spatially variable extinction toward the GC, it will likely prove difficult to fully characterize many of the detected planets toward the GC due to uncertain source star properties, distances, and other methods typically used to constrain planet and host star characteristics \cite{shvartzvald:2018a}. Nevertheless, all of the microlensing yields from the GC effectively act as a surplus to the overall GBTDS microlensing yield and should not be considered as part of the fixed fraction of microlensing planets that will yield mass measurements as originally set by the administration's requirements on microlensing exoplanets in the GBTDS.

Ultimately, adding a GC field to the GBTDS may have both positive and negative impacts on the overall microlensing exoplanet yields. This depends strongly on the final slew/settle times for the program. With current slew/settle times\footnote{https://roman.gsfc.nasa.gov/science/RRI/ROMAN\_Observatory\_Reference\_Information\_20210301.pdf}, and assuming the GBTDS fields of \cite{penny:2019a}, the addition of the high-cadence GC field would effectively lower the overall cadence for each field from ${\sim}15$ minutes to ${\sim}19$ minutes. Using equations 12, 13, and Table 6 of \cite{penny:2019a}, the number of detectable Earth-mass planets for this cadence is approximately $N_{\bigoplus, 19} = 0.91N_{\bigoplus, 15}$. In other words, ${\sim}20$ Earth-mass planets would be missed using the assumed parameters of \cite{penny:2019a} with no optimized slew/settle times. This assumes there are no Earth-mass planets detected in the GC survey that would balance this loss. If it is determined that there is extra efficiency that can be gained with faster slew/settle times (e.g. no diminished overall cadence), then adding a field to the GC is worth strongly considering. Finally, using the same assumptions as above (current slew/settle + current GBTDS fields of \cite{penny:2019a}), the addition of a low-cadence GC field would have a negligible effect on the overall cadence for each field (e.g. from ${\sim}15$ minutes to ${\sim}15.4$ minutes). We note there are several other possibilities for a lower cadence GC observing strategy, beyond the stated `minimal' strategy of 12 hour cadence. One additional example would be a high cadence GC survey during only the first and last seasons, and a lower cadence GC survey conducted during the middle seasons. While these 'mixed' cadence strategies do not optimize the yield for the GC-related science described here, they may represent a reasonable compromise while maintaining the level-1 requirements on exoplanets for the GBTDS.

\subsection{Monitoring SgrA*} \label{sec:sgra*-monitoring}
The gravitational influence of SgrA* can be seen through monitoring of early-type and late-type stars orbiting the central potential \cite{paumard:2006a, do:2013a}, as well as gas and dust streams orbiting at distances up to 1pc from SgrA* \cite{morris:1996a, genzel:2010a}. An intriguing filamentary gas feature, named X7, was recently reported to be undergoing fragmentation as it approaches SgrA* \cite{ciurlo:2023a}. This is one recent example of intriguing dynamical interactions between SgrA* and its immediate surroundings that are reachable by \textit{Roman}.

SgrA* itself has a near-IR counterpart that can exhibit flares lasting ${\sim}$1 hour \cite{genzel:2003a}. During these flares, the source flux can vary by a factor of 100 or more \cite{ghez:2004a, genzel:2003a}. The duration and intensity of these flares means that the optimal \textit{Roman GB+GC} observing strategy can easily sample these events. This high cadence monitoring will be extremely helpful for distinguishing different emission models that have traditionally been used to describe SgrA*'s quiescent phase. A long-standing mystery that \textit{Roman GB+GC} could address is the interpretation of SgrA*'s extremely low accretion rate that prior multi-wavelength studies have deduced. The minimal observing strategy of Table \ref{tab:observing-strat} may result in these short timescale flaring events being missed by the survey.

Finally, with either the minimal or optimal observing strategy presented here, long timescale evolution and other trends at the location of SgrA* can be studied. Having continuous, long-term tracking of the source and its emission (in near-IR) will aid in understanding the specific quiescent nature of the SMBH.

\section{Proper Motions and Large-scale Dynamics}\label{subsec:pms-clusters}
The recommended \textit{Roman GB+GC} survey would measure high-precision
stellar proper motions over an unprecedented area of the GC,
providing at least 50\% azimuthal coverage out
to R $\sim$13.5' (33 pc) from SgrA* (Figure \ref{fig:combo_map_with_chips}).
Due to the high extinction and stellar crowding at the GC, 
it is very expensive to make comparable observations
from the ground (where adaptive-optics or speckle-holography 
techniques are required) or from other space-based platforms
(e.g., \textit{Roman} offers a FOV 100x larger than \textit{Hubble}). 
Thus, the \textit{Roman-GC} survey is strongly synergistic 
with other facilities as it provides large-scale 
dynamical context for the GC as a whole. 

Here we highlight three science cases enabled 
by this survey: 
the formation and structure of the NSC and NSD,
the mass distribution and gravitational potential of the GC,
and the lifecycle, kinematics, and IMF of 
star clusters born in the region.
We also discuss the predicted astrometric performance 
of the survey in more detail.

\subsection{The NSC and NSD: Key Components of Galactic Nuclei}
\label{sec:nsc_nsd}
Nuclear star clusters and nuclear stellar disks are commonly 
observed in the central regions of spiral galaxies \cite{gadotti:2020a, neumayer:2020a},
and are key components for understanding the evolution
of galactic nuclei.
However, it is not yet clear how these features form or how they relate to one another.
In the Milky Way, the distinction between the NSC
and NSD is primarily based on the 
observed density profile of stars \cite[e.g.,][]{Chatzopoulos:2015lq, Gallego-Cano:2020fx}, 
and there is growing evidence that they may 
have different star formation histories \cite{Nogueras-Lara:2021ln, Chen:2023os,Schodel:2023fm}. 
Measuring the kinematics of these 
structures will reveal the relationship between
their formation mechanisms, 
as well as their connection to gas flows 
along the galactic bar.
Past efforts to model the kinematics of the NSC
and NSD have been limited by small sample 
sizes ($\lesssim$10,000 stars) and/or 
limited azimuthal coverage \cite[e.g.,][]{Do:2013hi, Chatzopoulos:2015lq, Feldmeier-Krause:2020yk, Sormani:2020my, Libralato:2021ck, Sormani:2022lz, Nogueras-Lara:2022cz}, 
due to the difficulty of measuring
stellar proper motions over large regions.

\textbf{The \textit{Roman GB+GC} survey overcomes these challenges
by measuring proper motions for $\sim$3.3 million stars over a region at least
$\sim$7x larger than previous proper motion studies}.
This represents a monumental gain in terms of both
radial coverage and sample size. 
Reaching R = 13.5', this survey extends well into the regime
where the NSD dominates the 
stellar density profile, and thus it becomes 
possible to cleanly model the two components (Figure \ref{fig:nsc_nsd}).

\subsection{The Mass Distribution and Gravitational Potential of the GC}
\label{sec:mass}
The \textit{Roman GB+GC} survey will
also significantly improve constraints on the mass distribution and gravitational 
potential of the GC.
The NSC and NSD dominate the gravitational potential 
of the galaxy between 1 pc $\lesssim$ R $\lesssim$ 300 pc \cite[e.g.,][]{Launhardt:2002hl},
but since their mass distributions are not well known,
there are significant uncertainties in the gravitational potential \cite{Sormani:2020my}.
This has many implications for a range of physical processes
in the region, as the gravitational potential plays a 
major role in the transportation of material to the GC from the galactic bar \cite[e.g.,][]{Tress:2020zj},
the formation of the observed ring-like structure of dense gas \cite[e.g.,][]{Molinari:2011dq},
the orbits of star clusters like Arches and Quintuplet \cite{hosek:2022a},
and may trigger star formation in dense molecular clouds in the region \cite[e.g.,][]{Kruijssen:2015mv, Kruijssen:2019kx}.

The recommended survey will greatly improve measurements of the 
mass distribution of the NSC and NSD, not only due 
to the large radial coverage, but also because
of the high azimuthal completeness it provides
over the survey region. 
{\bf This allows for unbiased measurements of the 
shape, rotation, and velocity anisotropy of the 
NSC and NSD, properties which would otherwise
be assumed or approximated when deriving the mass distributions
of these populations via kinematic modeling} \cite[e.g.,][]{Chatzopoulos:2015lq, Sormani:2020my}.

\subsection{The Lifecycle, Kinematics, and IMF of Young Star Clusters in the GC}
\label{sec:clusters}
The Arches and Quintuplet clusters are two of the most massive young clusters in the Galaxy. 
However, due to the strong tidal shear at the GC, these clusters are predicted to dissolve 
on timescales of $\lesssim20$ Myr \cite{kim:2000a, zwart:2002a}.
As a result, they are expected to exhibit 
large tidal tails that extend some 30$-$70 pc (12'$-$30') 
from the clusters \cite[][]{Habibi:2014qf}.
However, these structures have not yet been observed
due to two main challenges:
proper motions must be used to detect such tidal tails,
as differential extinction makes it impossible to identify 
young stars in the region through photometry alone \cite[e.g.,][]{Stolte:2008qy, Hosek:2015cs},
and the tails are expected to be composed of primarily 
lower-mass stars \cite[M $\lesssim$ 2.5 M$_{\odot}$;][]{Park:2018fj}.
{\bf The spatial coverage, astrometric accuracy, and sensitivity of the recommended
\textit{Roman GB+GC} survey would offer the first opportunity to 
detect and characterize the tidal tails of these clusters,
testing models of cluster evolution near the GC.}
This will reveal whether tidal disruption can resolve 
the apparent deficit of young clusters
near the GC relative to the recent star formation history \cite[the ``missing cluster problem''; e.g.,][]{Nogueras-Lara:2022rt},
or if additional modes of star formation other than star cluster formation
must be invoked.

In addition, characterizing the tidal tails of Arches and Quintuplet is required
to properly measure their IMFs.
The IMF is a key observational benchmark
for star formation theory, as its properties are tied to
the underlying physics driving the process \cite[e.g.,][]{Krumholz:2014ne, Offner:2014vn}.
For the Arches cluster, the IMF
has been found to be significantly different 
than those of star clusters in the Galactic disk for M $\gtrsim$ 2 M$_{\odot}$,
possibly due to the impact of the extreme 
GC environment on star formation \cite[e.g.,][]{Hosek:2019kk}.
However, several different IMF models are allowed, 
and extending the measurement to lower masses is required
to determine which model (and associated physical interpretation) is correct.
The loss of stars via tidal tails is expected to become much more significant
at such lower masses, and so characterizing the tidal tails 
is required to infer the IMF from the
present-day cluster population. 

Finally, the high astrometric precision of the \textit{Roman GB+GC}
survey will allow us to probe the internal kinematics
of the Arches and Quintuplet clusters.
{\bf With a proper motion accuracy of 15 - 25 $\mu$as/yr
from the minimal survey, 
it becomes possible to measure the velocity dispersion
profiles and thus dynamical mass of the clusters} \cite[e.g.,][]{Clarkson:2012ty}.
This allows for constraints on the mass function 
at stellar masses below what is possible to observe with current telescopes.
 
 \subsection{Astrometric Performance of the Survey}
 \label{sec:astrometry}
 To calculate the expected proper motion performance of the 
 \textit{Roman GB+GC} survey, we assume that all 
 astrometry will be extracted from the F146 images (FWHM = 0.105"),
 each of which have an exposure time of ${\sim}$50 s. 
 Based on the Roman WFI Imaging Sensitivity Calculator\footnote{https://roman.gsfc.nasa.gov/science/WFI\_technical.html; calculated June 2023},
 we expect to reach a signal-to-noise ratio (SNR) of $\sim$~20 -- 30 per image for 
 the stars at the faint end of our sample (F146 = 23 -- 24 VEGAMAG).
 We define the relationship between SNR and astrometric 
 error ($\sigma_{ast}$) as:
 
 \begin{equation}
 \sigma_{ast} = \alpha * \frac{FWHM}{SNR}
 \end{equation}
 
 \noindent where $\alpha$ is an empirical constant.
 We estimate $\alpha$ = 0.3  
 based on HST WFC3-IR F153M observations of the Arches and Quintuplet clusters 
 \cite{Hosek:2019kk, Rui:2019ae, hosek:2022a},
 which we expect to perform similarly to the recommended observations. 
 Thus, we predict $\sigma_{ast}$ = 1 -- 1.6 mas per image at the faint 
 end of the sample (Table \ref{tab:observing-strat}).
 
 For the purposes of astrometry, we can improve the astrometric
 performance by stacking multiple F146 images per astrometric epoch. 
 Since the Point Spread Function (PSF) of stars will begin to blur
 in stacked images due to their intrinsic motion,
 the timescale over which we can stack images with minimal
impact on the astrometric performance is constrained by the
 fastest moving stars in the sample.
 For the \textit{Roman GB+GC} survey, the fastest stars of interest
 would be hypervelocity stars produced via dynamical interactions 
 with SgrA* \cite{Hills:1988kq}, which may have ejection
 velocities upward of $\sim$1800 km/s \cite[$\sim$47.5 mas/yr;][]{Koposov:2020gz}.
We will stack F146 images until these stars move $\sim$1 mas 
(approximately the astrometric error of an individual image),
which means that each astrometric epoch will span 7 days.
Thus, for the minimal survey strategy (1 img / 12 hrs) 
and optimal observing strategy (1 img / 15 mins)
we will have 14 images and 672 images per epoch, respectively.

The astrometric performance per epoch ($\sigma_{ast,tot}$) is then:
\begin{equation}
\label{eq:ast-tot}
\sigma_{ast,tot} \approx \frac{\sigma_{ast}}{\sqrt{N_{img}}}
\end{equation}
where $N_{img}$ is the number of images per epoch \cite{WFIRST-Astrometry-Working-Group:2019lj}.
This yields $\sigma_{ast,tot}$ $\sim$ 0.3 -- 0.4 mas per epoch for the minimal survey
and $\sigma_{ast,tot}$  $\sim$ 0.04 -- 0.05 mas per epoch for the optimal survey.
The final proper motion precision of the survey ($\sigma_{pm}$) can then be estimated as:
\begin{equation}
\label{eq:pm}
\sigma_{pm} \approx \frac{\sigma_{ast,tot}}{\sqrt{\sum_{i=0}^{N_{ep}} (t_i - t_{ave})^2}}
\end{equation}
where $N_{ep}$ is the total number of epochs, $t_{ave}$ is the average time over all epochs, 
and $t_i$ is the time of the $i$th epoch. 
Assuming each observing season is 70 days long (and thus 10 epochs per season) 
and are spread over 5 years, then {\bf we anticipate a final 
proper motion precision of $\sigma_{pm}$ $\sim$ 15 -- 25 $\mu$as / yr ($\sim$ 1 km / s)
for the minimal survey and $\sigma_{pm}$ $\sim$ 2.5 -- 3.5 $\mu$as / yr ($\sim$ 0.15 km / s)
for the optimal survey.} 

If the observing seasons are distributed across the Fall and Spring
observing windows for the GC, then it becomes possible to measure
stellar distances via parallax for stars near the GC in the optimal survey.
At a distance precision of $\sim$200 pc, it becomes possible
to begin separating stars from in the NSD (R $\lesssim$ 200 pc
from SgrA*) from those in the Galactic bulge (R $\lesssim$ 1000 pc from SgrA*).
Assuming a typical distance of 8000 pc, this distance precision 
is achieved at a parallax uncertainty of approximately 3 $\mu$as.
The parallax uncertainty ($\sigma_{pi}$) can be
estimated as:
\begin{equation}
\sigma_{\pi} = \frac{FWHM}{\delta_\lambda * SNR_{tot}}
\end{equation}
where $SNR_{tot}$ is the total signal-to-noise across all images
and $\delta_{\lambda}$ is a factor defined from 0 -- 1 
that describes how well the parallactic ellipse is sampled
across the observations \cite{Lang:2009wi}. 
Assuming $\delta_{\lambda}$  = 0.8,
then $SNR_{tot}$ $\approx$ 44000 is required to achieve the 
target precision. 
Considering:
\begin{equation}
SNR_{tot} = \sqrt{\sum_{i = 0}^{N_{img}} (SNR_i)^2}
\end{equation}
where $SNR_i$ is the SNR achieved in the $i$th of $N_{img}$ total images.
Thus, for the optimal survey ($N_{img}$ = 40320), an 
$SNR_i$ $\approx$ 200 produces the necessary $SNR_{tot}$
for a meaningful parallax measurement.
{\bf This suggests that we will be able
to measure parallax-based distances to 
stars at the GC for F146 $\approx$ 19 mag,
which corresponds to high-mass stars (M $\gtrsim$ 8 M$_{\odot}$) in 
the young clusters and red clump stars in the older populations.}
We note that such parallax measurements will only be possible in the optimal survey;
for the minimal survey, which has much fewer images, 
$SNR_i$ $\approx$ 1500 is required
to achieve the required precision, which is not possible for the vast
majority of stars near the GC.


\section*{Appendix C}
\subsection*{Synergies with other White Papers} \label{sec:synergies}
The previous call for community input requested `science pitches' for each of the Roman CCS programs. This call received several pitches involving GC observations with Roman\footnote{https://rb.gy/vt2wt}, which demonstrates large and wide-ranging interest in visiting this region during the GBTDS. We note several interesting science pitches involving near-IR surveys from large ground-based observatories (PRIME, JASMINE) that will turn on soon and operate during the notional GBTDS mission. PRIME will offer high cadence photometric coverage using the same Roman chips and near-IR passbands, while JASMINE will conduct a near-IR astrometry mission starting in 2028. This astrometry program will measure very precise motions for mostly bright stars in the field ($H < 14$ mag), this information can act as a good calibration or systematics check for \textit{Roman} if \textit{Roman} acquires shallow images of these bright stars occasionally.

While it is uncertain how many of these science pitches will have accompanying white papers submitted for the current call, we are aware of at least one other Roman CCS white paper that specifically recommends GC observing strategies that synergize well with the studies presented in this work. The Bahramian CCS white paper \cite{bahramian:2023a} primarily focuses on black hole and neutron star science (e.g. XRBs and detached systems), and recommends moving one of the notional GBTDS fields to the GC and a second intervening field to connect the GC field with the other GBTDS fields (see Figure 1 of \cite{bahramian:2023a}). They show the total number of Chandra sources that can be monitored with the inclusion of their recommended fields is nearly $5\times$ higher than the number of sources covered by the notional GBTDS fields. We describe a similar science case in this white paper involving star-compact object binaries (e.g. detached systems) in Section 2 and Appendix A.

There are several synergies between our white paper and other GBTDS-related white papers \cite{Lam:2023a, street:2023a}. The Lam paper focuses on delivering high-quality astrometry from the data reduction pipelines and potentially increasing temporal coverage in the seasonal GBTDS gaps with Roman or other facilities, which would be of benefit to the GC-connected science presented here. The Street paper focuses on coordinating the survey strategy of the GBTDS with the Rubin observatory, by filling the GBTDS seasonal gaps with Rubin photometric coverage. This is particularly useful for long duration microlensing events caused by compact object lenses like black holes and neutron stars. Figure 2 of \cite{street:2023a} shows that the very wide field of Rubin allows a deep drilling field placed on the notional GBTDS to also cover the GC (or at least partially). This would clearly benefit the \textit{Roman GB+GC} case for long duration microlensing events.

\medskip
\printbibliography

\end{document}